\documentclass[prl,twocolumn,superscriptaddress]{revtex4}

\usepackage{graphicx}
\newcommand{\ket}[1]{\ensuremath{\left|#1\right\rangle}}

\usepackage[normalem]{ulem} %need for strikethrough
\usepackage{color} %needed for coloring of \todo etc.

\begin{document}

%\title{Microwave coherent memory with erbium spin ensemble}
\title{Microwave multimode memory with an erbium spin ensemble}

\author{S.~Probst}
\affiliation{Physikalisches Institut, Karlsruhe Institute of Technology, D-76128
Karlsruhe, Germany}
%\affiliation{DFG-Center for Functional Nanostructures (CFN), D-76128 Karlsruhe, Germany}

\author{H.~Rotzinger}
\affiliation{Physikalisches Institut, Karlsruhe Institute of Technology, D-76128
Karlsruhe, Germany}

\author{A.~V.~Ustinov}
\affiliation{Physikalisches Institut, Karlsruhe Institute of Technology, D-76128
Karlsruhe, Germany}

\author{P.~A.~Bushev}
\affiliation{Experimentalphysik, Universit\"at des Saarlandes, D-66123 Saarbr\"{u}cken, Germany}

\date{\today}

\begin{abstract}
Hybrid quantum system combining circuit QED with spin doped solids are an attractive platform for distributed quantum information processing. There, the magnetic ions serve as coherent memory elements and reversible conversion elements of microwave to optical qubits. Among many possible spin-doped solids, erbium ions offer the unique opportunity of a coherent conversion of microwave photons into the telecom C-band at $1.54\,\mu$m employed for long distance communication. In our work, we perform a time-resolved electron spin resonance study of an Er$^{3+}$:Y$_2$SiO$_5$ spin ensemble at milli-Kelvin temperatures and demonstrate multimode storage and retrieval of up to 16 coherent microwave pulses. The memory efficiency is measured to be 10$^{-4}$ at the coherence time of $T_2=5.6\,\mu$s. We observe a saturation of the spin coherence time below 50mK due to full polarization of the surrounding electronic spin bath.
\end{abstract}
\maketitle

%\pacs{42.50.Fx, 76.30.Kg, 03.67.Hk, 03.67.Lx, 76.30.-v}
%\keywords{Cooperative phenomena in quantum optical systems, Superconducting qubits, Quantum communication, Quantum computation architectures and implementations, EPR in condensed matter}

A future quantum communication technology will combine three basic types of subsystems: Transmission channels, repeater stations and information processing nodes~\cite{DiVincenzo2000, Gisin2007}. Similarly to classical communication networks, photons propagating through optical fiber channels are ideal for carrying quantum states over long distances and distribute entanglement between information processing nodes~\cite{Northup2014}. These computational nodes or quantum processors may be realized by employing single atomic or macroscopic solid-state systems. Among a plethora of solid state devices, superconducting (SC) qubits are one of the most promising building blocks for a future quantum computer~\cite{Yale2013}. Many ground breaking experiments have recently been demonstrated with SC qubits including the measurement of long coherence and relaxation times of up to 0.1\,ms~\cite{Paik2011}, coherent operation of up to three-qubit processors~\cite{Reed2012}, the implementation of a deterministic two-qubit gate~\cite{Steffen2013} and the realization of a basic surface code for fault tolerant computing~\cite{Barends2014}. These qubits operate at microwave frequencies and cryogenic temperatures. In order to embed them into the emerging quantum optical internet technology a coherent interface between optical and microwave photons is required~\cite{Tian2004}.

Ensembles of rare-earth (RE) ions doped into a crystal are a suitable system for coherent photon conversion between optical and microwave frequency bands~\cite{OBrien2014, Williamson2014, Twamley2014}. Such RE doped crystals are currently at the forefront of quantum communication research, where many thrilling achievements such as the demonstration of a quantum memory at the optical telecom C-band~\cite{Lauritzen2010}, high efficiency storage of optical photons~\cite{Hedges2010}, generation of entanglement between two RE doped crystals~\cite{Usmani2012} and quantum teleportation between a telecom O-band photon ($1.34\,\mu$m) and a RE doped crystal~\cite{Bussieres2014} have been reported. Also, RE doped crystals are considered to have a great potential as a multimode optical memory element in a future quantum repeater technology~\cite{Simon2007}, and storage and retrieval of 64 temporal optical modes at the single photon level has been demonstrated~\cite{Usmani2010}.

A crucial step towards the development of a microwave to optical interface requires highly efficient reversible mapping of temporal microwave modes into the rare-earth spin ensemble at a power level corresponding to a single microwave photon~\cite{Afzelius2013}. In that respect, circuit Quantum Electrodynamics (QED) with RE doped crystals at the femtowatt power level and at milli-Kelvin temperatures presents a first step towards the implementation of a quantum memory~\cite{Bushev2011, Staudt2012}. A strong coupling regime accompanied with large collective coupling strength 30-200\,MHz has been recently demonstrated~\cite{Probst2013, Tkalcec2014, Farr2014}. However, reversible mapping of temporal microwave modes in a RE spin ensemble has not been shown. So far, most of the time resolved microwave experiments with RE doped crystals have been limited by studying of their coherence and magnetic properties performed in conventional electron spin resonance~(ESR) spectrometers at temperatures above 2\,K~\cite{Bertaina2007, Bertaina2009, Baibekov2009, Baibekov2014}. In this work, we investigate the spin coherence properties of an Er$^{3+}$:Y$_2$SiO$_5$ (Er:YSO) crystal at milli-Kelvin temperatures and demonstrate the successful storage and on-demand retrieval of 16 weak coherent microwave pulses.

\begin{figure}[ht!]
\includegraphics[width=0.8\columnwidth]{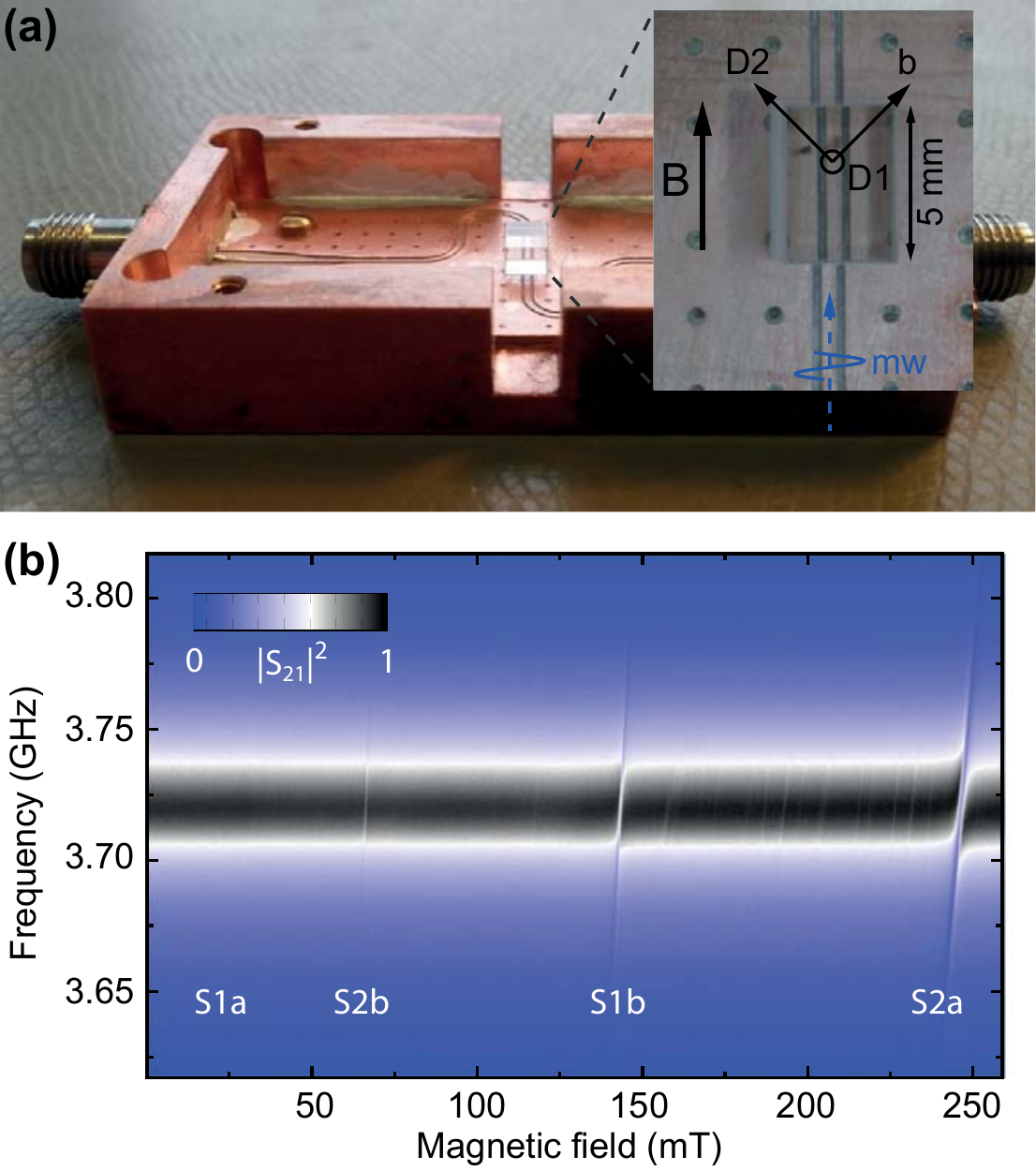}
\caption{(Color online) \textbf{(a)} Experimental setup: The Er:YSO crystal is placed on a $\lambda/2$ copper coplanar waveguide resonator. Inset: Orientation of the crystal with respect to the DC magnetic field. \textbf{(b)} Transmission ESR spectrum of Er:YSO coupled to the resonator. The color code on the right side illustrates the magnitude of microwave transmission $|S_{21}(\omega)|$ through the chip.}\label{Setup}
\end{figure}

Figure~\ref{Setup}(a) shows a picture of the experimental setup. We investigate a single Er:YSO crystal (Scientific Materials Inc.) doped with 0.005\% atomic concentration of Er$^{3+}$ ions and with dimensions of 3 x 4 x 5\,mm. The inset of this figure presents the orientation of the crystal axes $D_1$, $D_2$, $b$ with respect to the applied DC magnetic field $B$. The crystal orientation ($\theta\simeq 45^{\circ}$) maximizes the coupling strength for the high-field transition $S_{2a}$~\cite{Probst2013}. The crystal is placed on top of a coplanar waveguide $\lambda/2$ microwave resonator fabricated on high-frequency laminate Rogers TMM 10i. In contrast to our previous investigations~\cite{Bushev2011, Probst2013, Tkalcec2014}, a non-superconducting copper resonator is employed. Such a resonator does not perturb the magnetic DC field, therefore, we expect to attain a minimal inhomogeneous spin linewidth of the erbium spin ensemble. The width of the coplanar waveguide is about 0.5\,mm with gaps of about 0.25\,mm. The comparably large geometric dimensions of the resonator are beneficial for optical access and for attaining an intrinsic quality factor of $Q_i\simeq 400$. The experiment is placed inside a copper housing and cooled by a BlueFors LD-250 dilution fridge to 25\,mK. For further details on the experimental setup, we refer to Ref.~\cite{ProbstPhD}.

Initially, the sample is characterized in the common way by continuous wave (cw) microwave transmission spectroscopy while sweeping the magnetic field~\cite{Shuster2010,Tobar2013}. Figure~\ref{Setup}(b) presents the resulting spectrum recorded at 25\,mK and a probing power of $\sim100$\,fW. The transmitted amplitude is color coded and all four sub-ensembles with g-factors 14.2, 4.0, 1.9 and 1.1 are resolved. The electronic transition close to 140\,mT shows a weak anticrossing indicating the onset of strong coupling. A clear normal mode splitting is observed for the high field transition $S_{2a}$ at a magnetic field of 246\,mT.

In this work, we focus our analysis on the high field transition $S_{2a}$ at 246\,mT with $\textrm{g}=1.1$. The avoided level crossing is analyzed by first fitting the resonator far away from the transition. This yields a resonance frequency of $\omega_0/2\pi = 3.721$\,GHz and a HWHM linewidth of $\kappa/2\pi = 8.2\pm 0.1$\,MHz. By using this parameters, we extract a collective coupling strength of $v_N/2\pi = 13.2\pm 0.7$\,MHz and a HWHM inhomogeneous spin linewidth of $\Gamma^\star_2/2\pi = 7.3\pm 0.4$\,MHz. %We note here, that minimal spin linewidth measured with the present copper resonator was $\Gamma^\star_2/2\pi = 3.5\pm 0.4$\,MHz.

\begin{figure}[ht!]
\includegraphics[width=1\columnwidth]{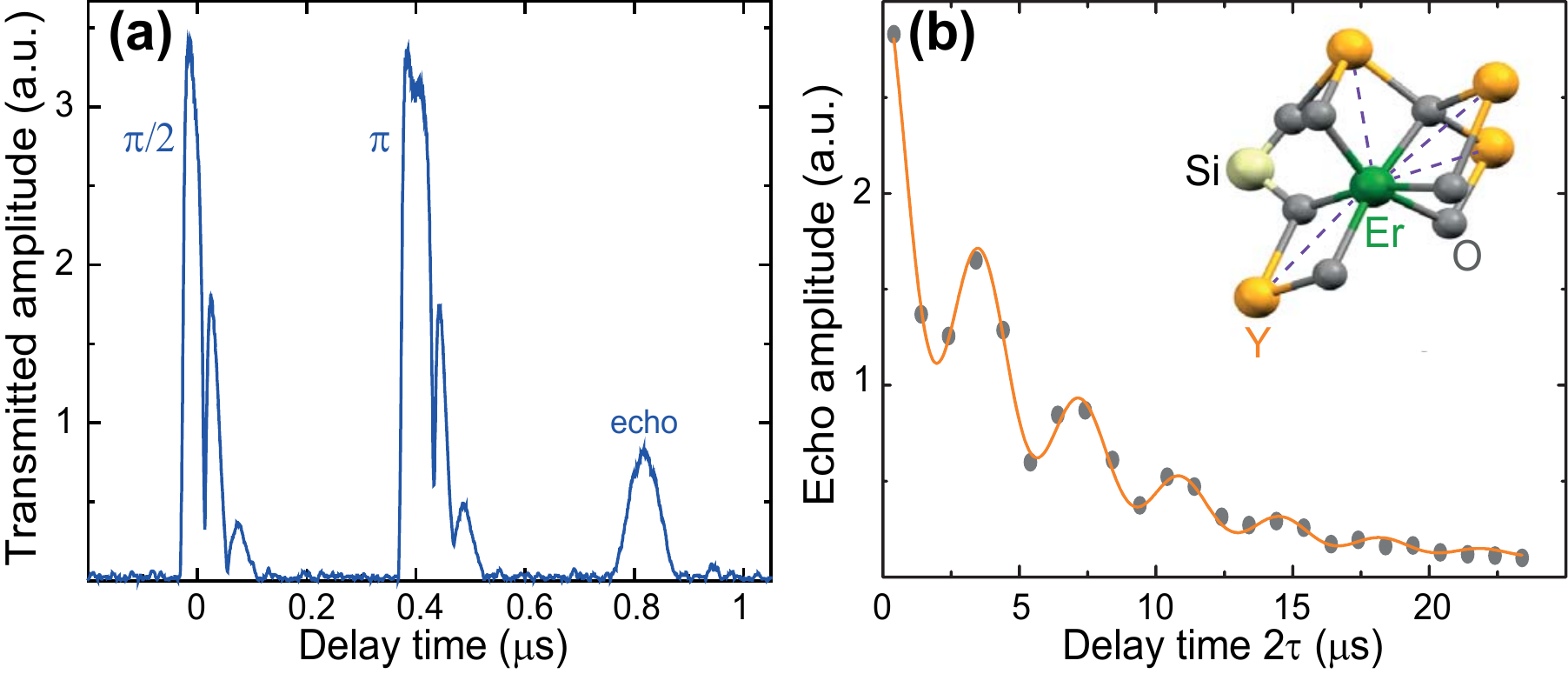}
\caption{(Color online) \textbf{(a)} Hahn echo sequence with a pulse separation of $\tau=400\,$ns. Each pulse is accompanied by vacuum Rabi oscillations. The $\pi/2$-pulse has a width of 20\,ns. The spin echo appears as a 100\,ns wide pulse due to the finite bandwidth of the resonator. \textbf{(b)} The echo decay is modulated by prominent oscillations, which are known as electron spin echo envelope modulation (ESEEM). This originates from the coupling of the erbium electronic spins to the nuclear spins of the neighboring $^{89}$Y ions. The solid line presents a fit to the data yielding a $T_2\simeq 5.6\,\mu$s. }\label{Echo}
\end{figure}

On the basis of the cw spectroscopy, we performed time resolved ESR experiments in the temperature range 0.02-1\,K. Due to the large mode volume of the resonator compared to the previous experiments with SC resonators~\cite{Bushev2011, Probst2013, Tkalcec2014}, strong microwave pulses (1-10\,mW) are necessary for coherent spin manipulation. Therefore, no additional cryogenic attenuation was used in the experiment, see Ref.~\cite{ProbstPhD} for details. Thus, we estimate a the number of thermal photons in the cavity to be about $n_{\text{ph}}\sim 50$. For comparison, the number of spins coupled to a resonator mode is $N_s\sim3\times 10^{13}\gg n_{\text{ph}}$, therefore, the thermal photons do not affect the polarization of the spin ensemble significantly.

Figure~\ref{Echo}(a) shows a two pulse Hahn echo (2PE) sequence $\pi/2-\tau-\pi$ followed by an echo emission. The length of $\pi$ pulse is set to be 40\,ns and its magnitude is about 10\,mW.  The transmitted pulse shapes of the $\pi/2$ and $\pi$-pulse feature vacuum Rabi oscillations. Thus, a short but strong pulse does not affect the Tavis-Cummings dynamics of the hybrid system. Note that the free induction decay is not observable here because the large inhomogeneous broadening results in a very short pure dephasing time $T_2^*\simeq 22\,$ns.

Figure~\ref{Echo}(b) presents the amplitude of the echo signal in dependence on the total delay time 2$\tau$. The oscillatory modulation of the echo decay is attributed to electron spin echo envelope modulation (ESEEM)~\cite{PulsedESR}, which originates from the dipole-dipole coupling of the erbium electronic spin to nuclear spins in close proximity~\cite{Guillot2007}. The closest magnetic dipoles are the nuclear spins of the yttrium $^{89}$Y ions with magnetic moment $\mu_Y=-0.275\mu_N$ at an average distance of $\sim4\,$\AA. The fit to the modulated exponential decay yields a spin coherence time of $T_2=5.6\,\mu$s. We would like to point out here that conventional ESR experiments with Er:YSO at 6\,K show a spin coherence time $T_2\sim 0.3\, \mu$s.

%Temperature dependence

In the following, we study the decoherence mechanisms of the erbium spin ensemble $S_{2a}$. There are two major contributions to decoherence: (I)~Dephasing induced by the surrounding magnetic environment, i.e. other electronic or nuclear spins~\cite{Takahashi2008}, and (II)~direct spin-spin interaction within the same spin ensemble~\cite{Tyryshkin2011}. Source (I) is addressed by investigating the temperature dependence of the coherence time $T_2$. The increase of the coherence time due to the "freezing" of surrounding spin bath has been observed in mm-wave ESR with paramagnetic materials~\cite{Kutter1995,Takahashi2008}. Those experiments were performed above 2\,K requiring a magnetic field of several Tesla to ensure full polarization of the electronic spin bath. In contrast, the base temperature of our experiment (25\,mK) entails operation at the GHz frequency range while maintaining full polarization at only $\sim100$\,mT, see also Fig.~\ref{Setup}(b).

\begin{figure}[ht!]
\includegraphics[width=0.9\columnwidth]{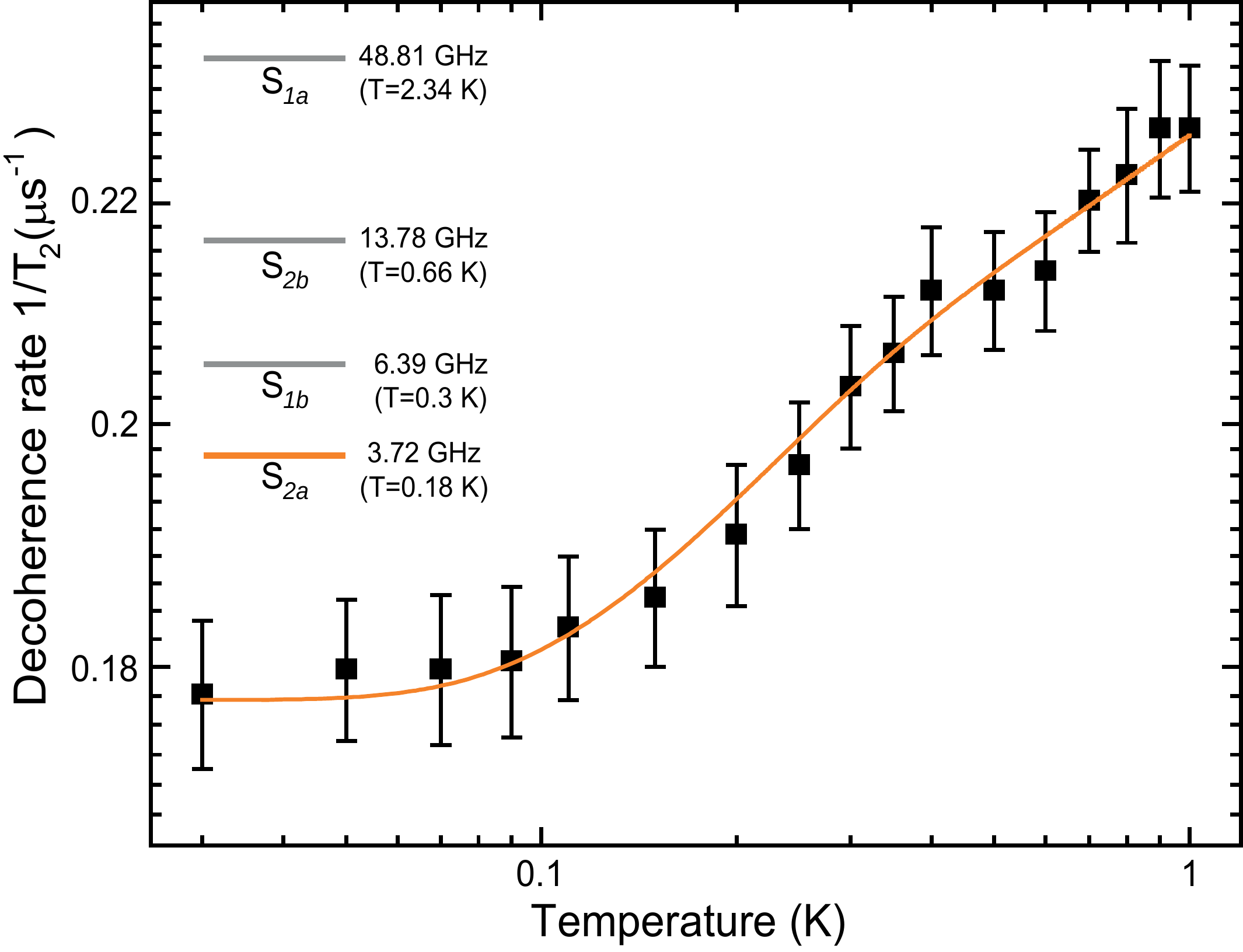}
\caption{(Color online) Temperature dependence of the decoherence rate $1/T_2$ of $S_{2a}$ ensemble measured by the spin-echo experiment. Note the semi-logarithmic scale. Between 30\,mK and 1\,K, the $T_2$ time decreases from approximately 5.6\,$\mu$s to 4.4\,$\mu$s. The solid line is a fit to theory, see Eq.~(\ref{T2vsT}), which takes into account thermal fluctuations from the surrounding erbium sub-ensembles. Inset: Transition frequencies and effective temperatures at 246\,mT of all sub-ensembles.}\label{T-dependence}
\end{figure}

Figure~\ref{T-dependence} presents the temperature dependence of Er:YSO spin coherence time $T_2$ from 30\,mK to 1 K. In the displayed temperature range, the $T_2$ time increases by approximately 1\,$\mu$s. Below 50\,mK, $T_2$ remains constant, indicating that magnetic fluctuations are "frozen out"~\cite{Takahashi2008}. Since the spin-lattice relaxation time is measured to be $\sim 10$\,s in the experimental temperature range, it does not limit $T_2$.

In order to account for the observed behavior of $T_2$, we have to consider all four sub-ensembles of erbium spin in YSO crystal. The spin concentration of each ensemble is $n_s\simeq 10^{17}\,$cm$^{-3}$ and their g-factors vary from 1.9 to 14.2. These ensembles present a considerable fluctuating magnetic background. To put the Zeeman transition frequencies into perspective, Fig.~\ref{T-dependence} also provides the effective Zeeman temperatures of each sub-ensemble $T_i=\textrm{g}_i \mu_B B/k_B$, where $k_B$ is the Boltzmann constant. As the temperature rises, the background spins perform random flip-flops altering the precession frequencies of the S$_{2a}$ ensemble~\cite{Klauder1962}. The occupation probabilities are given by the Boltzmann statistics, $P_{\uparrow} = [1+\exp (T_i/T)]^{-1}$ and $P_{\downarrow} = [1+\exp (-T_i/T)]^{-1}$. By including a residual dephasing rate $\Gamma_r$, the temperature dependence of $T_2$ is described by~\cite{Kutter1995,Takahashi2008}

\begin{equation}\label{T2vsT}
\frac{1}{T_2 (T)} = \Gamma_r+ \sum_{i=1}^{3} {\frac{\xi}{\left( 1+e^{T_i/T} \right) \left( 1+e^{-T_i/T} \right)}},
\end{equation}
where $\xi$ is a temperature independent free parameter.

The solid line in Fig.~\ref{T-dependence} presents a fit of Eq.~(\ref{T2vsT}) to the data, where $\Gamma_r$ and $\xi$ are the only free parameters. Below 100\,mK, the temperature dependence saturates and the fit yields a $\Gamma_r = (5.63\,\mu$s$)^{-1}$. Also, the fit reproduces the data including the slightly shallower slope towards larger temperatures. Thus, the spin flip-flop processes of the other three sub-ensembles dominate the temperature dependence of coherence time $T_2$.

The second source of decoherence are spin flip-flop processes within the same sub-ensemble. These give rise to instantaneous spin diffusion, which cannot be refocused by a standard 2PE sequence. However, it is possible to measure this dipolar interaction with a modified 2PE sequence where the angle of the second pulse $\theta_2$ is varied from $0$ to $\pi$~\cite{Brown2011}. The $\pi$ angle corresponds to a standard 2PE sequence and refocuses the magnetic field inhomogeneity and low frequency magnetic noise. If a spin flip-flop occurs, the $\theta_2$-pulse cannot refocus this interaction because the pulse flips both spins involved in the interaction as they belong to the same spin ensemble. The longest coherence time measured by applying the $\pi/2-\tau-\theta_2$ sequence is $T_2=7$\,$\mu$s. From the dependence $T_2(\theta_2)$ we infer a dipole-dipole coupling strength between neighbouring spins in $S_{2a}$ of $v_D/2\pi=12$\,kHz.

%Multimode memory part

Next, we demonstrate a microwave multimode memory with Er$^{3+}$:Y$_2$SiO$_5$ by storing and retrieving 16 weak coherent pulses. In a multimode spin ensemble quantum memory, an incoming photon is mapped onto a coherent spin wave of the ensemble~\cite{Wesenberg2009}. A single photon being absorbed by an ensemble of $N$ spins manifests itself as a coherent superposition of all possibilities of one spin being excited with the rest in the ground state. In an inhomogeneously broadened ensemble, one has to consider the different precession frequencies of the spins, yielding
\begin{equation}
\ket{\Psi}_{\text{ph}} = \frac{1}{\sqrt{N}} \sum_{k=1}^N{ \ket{\downarrow_1 \downarrow_2 \cdots \uparrow_k \cdots \downarrow_N} e^{-i\delta_k t }}\,,
\end{equation}
where $\delta_k$ denotes the detuning of the $k$-th spin from the mean precession frequency of the ensemble. As the time $t$ progresses, this state dephases into a dark mode, such that the bright mode can absorb the next photon. A desired but feasible quantum memory would be able to store multiple photons. In a spin ensemble, the number of stored modes is limited by $N_m=T_2/T_2^*$. However, in our experiment, the finite bandwidth of the resonator stretches the echo emission to about 100\,ns reducing the number of effective modes to approximately 56.
%therefore $N_m=T_2 \kappa \approx 100$.
Upon application of a $\pi$-pulse, the time evolution is reversed and all the dark modes rephase again emitting a photon. Thus, the time order of the incident photons is reversed with respect to the sequence of emitted photons.

\begin{figure}[ht!]
\includegraphics[width=0.9\columnwidth]{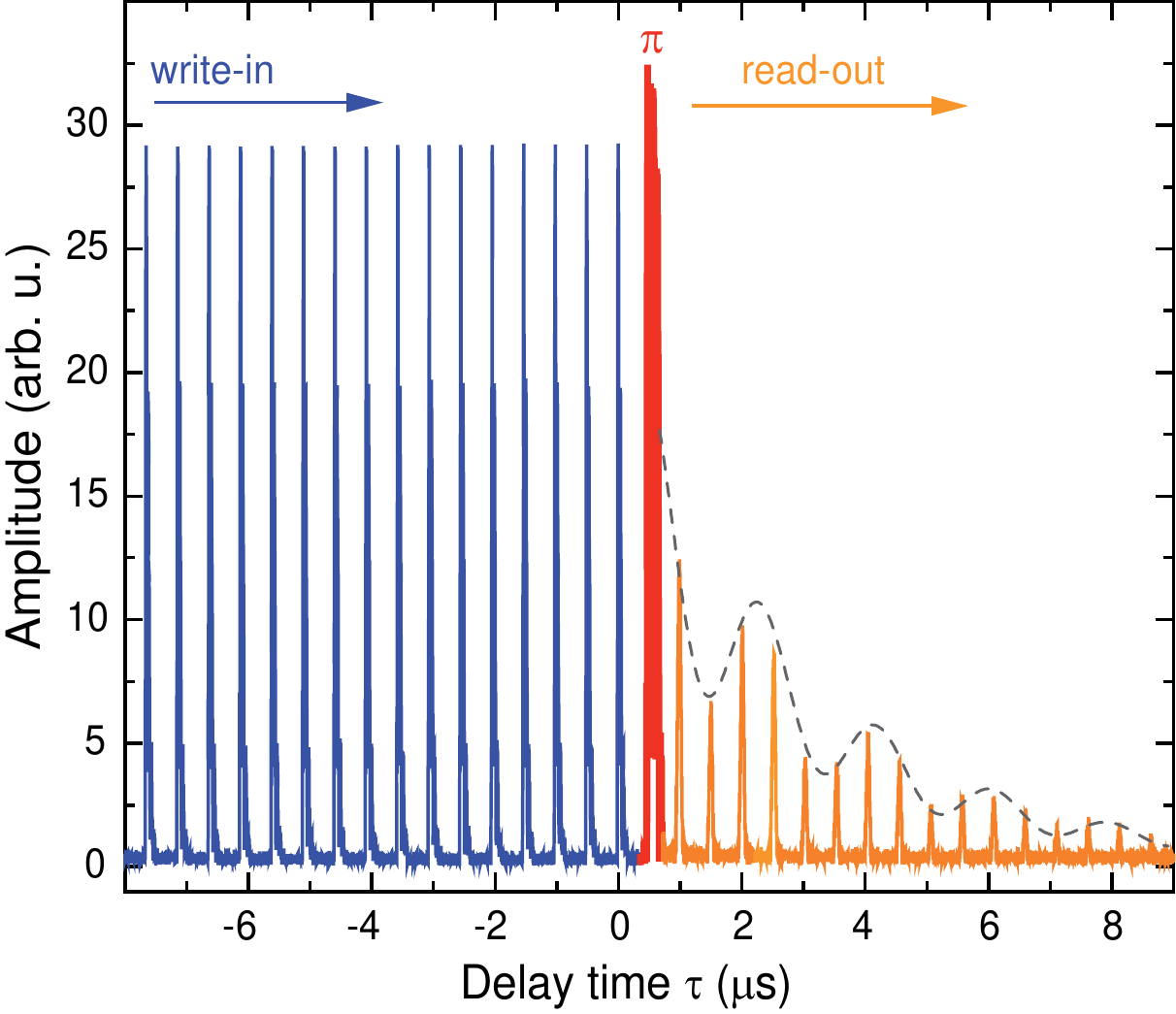}
\caption{(Color online) Storage of 16 microwave coherent pulses in the $S_{2a}$ erbium spin ensemble. All input pulses are of the same height and 10\,ns width. The amplitudes of the refocusing and input pulses are in the saturation limit of our amplifiers and appear smaller in the plot. The retrieved pulse stream shown in the plot is exponentially damped and exhibits oscillatory behavior due to the coupling of the electronic spins to the nuclear spins of $^{89}$Y ions (ESEEM). The dotted curve is identical to the ESEEM signal presented in Fig.~\ref{Echo}(b). The ESEEM signal is superimposed in the present graph to demonstrate the oscillating envelope of the read-out pulses.}\label{Memory}
\end{figure}

Figure~\ref{Memory} presents the complete sequence of writing-in and reading-out of 16 microwave coherent pulses. Every incoming pulse contains about $\sim 4.6\times 10^{11}$ photons and corresponds to power level of 115 $\mu$W. The signal is averaged 595 times with a repetition rate of 10\,mHz. As expected, the emitted pulse train decays towards longer storage times. Interestingly, the decay pattern shows modulations attributed to ESEEM.

The performance of a memory is defined by the efficiency for a pulse being stored during the coherence time $T_2$ (5.6\,$\mu$s for the present system) of the spin ensemble. In order to determine the memory efficiency, the pulse energy at $T_2$ after the refocusing pulse is analyzed and compared to the energy of the input pulse. Note, the energy of the input pulses was determined in a separate calibration procedure. We obtain an energy retrieval efficiency of $10^{-4}$ at $T_2$.

Recently, storage of microwaves in the high power regime ($n_{\text{ph}}\sim 10^{14}$) has been reported, by using a conventional ESR spectrometer~\cite{Wu2010}. The memory efficiency in that experiment was $10^{-10}$. A more recent investigation by Grezes et al.~\cite{Grezes2014} demonstrates storage and retrieval of extremely weak microwave pulses ($n_{\text{ph}}\sim 3$ photons) in a nitrogen vacancy spin ensemble with an efficiency of $2\times 10 ^{-4}$ at temperature of 300\,mK.

According to Grezes et al.~\cite{Grezes2014}, the theoretical limitation for the efficiency is given by $T_2$. This should yield an efficiency of $10^{-3}$ for the current crystal. Therefore, one additional order of magnitude is lost by the experimental design. This may be improved by the application of optimal control pulse schemes~\cite{Sigillito2014} and cleverer resonator designs with very homogeneous AC fields~\cite{Benningshof2013} in conjunction with surface spin doped samples~\cite{Probst2014a}. Ultimately, RE ion doped crystals with better coherence properties are required. For instance, crystals predominantly doped with RE isotopes with non-zero nuclear spin ($^{167}$Er or $^{143}$Nd) features $T_2\simeq 100 \,\mu$s~\cite{Bertaina2007, Baibekov2014}, while maintaining sufficient optical depth~\cite{Baldit2010}. Even longer coherence times of the hyperfine transitions are attained by operating the microwave memory at a \emph{clock transition}~\cite{McAuslan2012, Wolfowicz2013}, where the spin's precession frequency is to first order insensitive to magnetic field fluctuations. Alternatively, one can transfer microwave excitations from electronic to nuclear spins of rare-earth ions~\cite{Wolfowicz2014}. Future progress towards the development of a microwave to optical quantum converter is only achieved with experiments on efficient storage of single microwave photons. Miniaturization of the resonator and the usage of parametric and cryogenic HEMT amplifiers will enable operation in such regime.

In conclusion, we have presented a detailed pulsed ESR study of an Er:YSO crystal at milli-Kelvin temperatures. The temperature dependence of the spin coherence time confirms full polarization of all electronic spin baths at low temperature. The clear ESEEM signal originating from the coupling of the erbium spins to $^{89}$Y nuclear spins holds a great potential of employing them as a nuclear spin quantum memory. The coherence time of the spin ensemble at milli-Kelvin temperature is measured to be about 5.6\,$\mu$s, which allowed us to store and retrieve up to 16 weak coherent pulses. The memory efficiency of $10^{-4}$ is close to state-of-the-art experiments. The presented work paves the way towards applying rare-earth doped crystals as a reversible conversion element between microwaves and telecom C-band photons at 1.54\,$\mu$m.

We thank John Morton for useful discussions and the critical reading of the manuscript. S.~P. acknowledges financial support by the LGF of Baden-W\"{u}rttemberg. This work was supported by the the BMBF Program “Quantum Communications” through the project QUIMP.

\bibliography{mwstorage_bib}

\end{document}